\documentclass{article_saj}
\usepackage{graphicx,saj,multicol,subeqnarray}
\usepackage{natbib}
\usepackage{float}
\usepackage{xcolor}
\usepackage{widetext}
\usepackage{url}
\usepackage{bm}
\usepackage{tikz} 
\usepackage{pifont} 
\usepackage{amsfonts}
\usepackage{amssymb}
\usepackage{amsmath,upgreek}
\usepackage{titlesec}
\titlelabel{\thetitle.\quad}
\definecolor{xlinkcolor}{cmyk}{1,0.6,0,0}
\usepackage[bookmarks=false,         
     pdfnewwindow=true,      
     colorlinks=true,    
     linkcolor=xlinkcolor,     
     citecolor=xlinkcolor,     
     filecolor=xlinkcolor,  
     urlcolor=xlinkcolor,      
final=true
]{hyperref}

\def\papertype{\ \hfill\ Original scientific paper}

\def\udc{52}
\DeclareTextCommand{\DJ}{OT1}{%
  \raisebox{-0.1ex}{\scalebox{0.75}[1.4]{--}}\kern-.4em D%
}

\begin{document}
\parindent=.5cm
\baselineskip=3.8truemm
\columnsep=.5truecm
\newenvironment{lefteqnarray}{\arraycolsep=0pt\begin{eqnarray}}
{\end{eqnarray}\protect\aftergroup\ignorespaces}
\newenvironment{lefteqnarray*}{\arraycolsep=0pt\begin{eqnarray*}}
{\end{eqnarray*}\protect\aftergroup\ignorespaces}
\newenvironment{leftsubeqnarray}{\arraycolsep=0pt\begin{subeqnarray}}
{\end{subeqnarray}\protect\aftergroup\ignorespaces}
%


\markboth{\eightrm YARKOVSKY/YORP EFFECTS IMPLEMENTATION} 
{\eightrm M. FENUCCI and B. NOVAKOVI\'C}

\begin{strip}

{\ }

\vskip-1cm

\publ

\type

{\ }


\title{\textsc{MERCURY} AND \textsc{ORBFIT} PACKAGES FOR NUMERICAL INTEGRATION OF PLANETARY SYSTEMS:  IMPLEMENTATION OF THE YARKOVSKY AND YORP EFFECTS}


\authors{Marco Fenucci$^{1}$, Bojan Novakovi\'c$^{1}$}

\vskip3mm


\address{$^1$Department of Astronomy, Faculty of Mathematics,
University of Belgrade\break Studentski trg 16, 11000 Belgrade,
Serbia}


\Email{marco\_fenucci@matf.bg.ac.rs, bojan@matf.bg.ac.rs.rs}


\dates{-}{-}


\summary{For studies of the long-term evolution of small Solar System objects, it is fundamental to add the Yarkovsky and Yarkovsky-O'Keefe-Radzievskii-Paddack (YORP) effects in the dynamical model. Still, implementations of these effects in publicly available $N$-body codes is often lacking, or the effects are implemented using significantly simplified models.
In this paper, we present an implementation of the coupled Yarkovsky/YORP effects in the \textnormal{\textsc{mercury}} and \textnormal{\textsc{orbfit}} $N$-body codes. Along with these two effects, we also included the effects of non-destructive collisions and rotationally induced breakups to model the asteroid spin state properly. Given the stochastic nature of many incorporated effects, the software is suitable for statistical dynamical studies. Here we explained the scientific aspect of the implementation, while technical details will be made freely available along with the source codes.  }



\end{strip}

\tenrm


\section{INTRODUCTION}
The dynamics of Solar System objects is determined by their mutual gravitational interactions. However, proper modeling of the dynamics of small Solar System objects, smaller than about 30-40~km in size, also requires non-gravitational forces to be included in the model. For asteroids, two relevant such effects are the Yarkovsky and Yarkovsky-O'Keefe-Radzievskii-Paddack (YORP). 
The Yarkovsky effect is caused by sunlight. When an asteroid heats up in the Sun, it eventually re-radiates the energy away as heat, creating a tiny thrust. This recoil acceleration is much weaker than the gravitational forces, but it can produce substantial orbital changes over long timescales. The same physical phenomenon is also responsible for the YORP effect, a thermal torque that, complemented by a torque produced by scattered sunlight, can modify the rotation rates and obliquities of small bodies as well \citep[see e.g.][]{bottke-etal_2006,vokrouhlicky-etal_2015}.
Therefore, the Yarkovsky effect changes the orbital motion of an asteroid, while the YORP effect changes the spin state. However, as the magnitude of the Yarkovsky effect depends on the spin state, the two effects are coupled. For this reason, a high-precision analysis of the orbit evolution of small objects requires both effects to be considered. 

It has been proven that the Yarkovsky and YORP effects play an important role in several phenomena in our Solar System, including the transport of objects from the asteroid belt to the near-Earth region \citep[e.g.][]{1999Sci...283.1507F,2002Icar..156..399B,2017A&A...598A..52G}, the spreading of asteroid families \citep[e.g.][]{2006Icar..182..118V,2015Icar..257..275S}, and evaluation of the impact risks \citep[e.g.][]{2019A&A...627L..11D,2021AJ....162..277R}.

Nowadays, numerical integrators are fundamental tools to study problems of celestial mechanics. In the context of our Solar System, long-term numerical simulations are often needed to understand and constrain the past dynamical evolution of planets, satellites, asteroids, and comets. 
Over the years, different integration schemes have been developed to propagate the gravitational $N$-body problem, and most of them have been included in software packages written in different programming languages. 
The three most widely used numerical integration packages in the planetary science community are \textsc{swift} \citep{1994Icar..108...18L}, \textsc{mercury} \citep{chambers-migliorini_1997}, and \textsc{orbit9} \citep{1988CeMec..43....1M} that is included in the \textsc{OrbFit}
package. In addition to these well established packages, \textsc{rebound} \citep{2012A&A...537A.128R} became a top-rated tool in the last few years. The last decade also saw a fast improvement in the performances of graphical processing units (GPUs). GPUs can efficiently handle parallel computing, and for this reason, packages such as \textsc{genga}\footnote{\url{https://bitbucket.org/sigrimm/genga}}  \citep{2014ApJ...796...23G} have been developed to exploit the advantages of this kind of hardware. Finally, new integrators are also under development \citep[e.g.][]{2022NewA...9001659Z}, answering the need for increased computational and modeling demands.


In the basic version of these integrators, stars and planets are treated as point masses, and their orbits evolve according to their mutual gravitational attraction. On the other hand, small bodies such as asteroids and comets are treated as massless particles, meaning that their dynamics is influenced by the gravitational attraction of stars and planets, but they do not cause any additional force on other bodies. Some of these integrators also include effects due to the oblateness of the central body, or relativistic effects. An example is the \textsc{mercury-t} extension of the \textsc{mercury} code, which takes into account tides, general relativity, and the effect of rotation-induced flattening in order to simulate the dynamical and tidal evolution of multi-planet systems \citep{2015A&A...583A.116B}.

Among the public distributions of $N$-body codes mentioned above, the Yarkovsky effect has been implemented in \textsc{swift}, \textsc{genga} and \textsc{orbit9}, while only \textsc{swift} has also the YORP effect implemented \citep{2011MNRAS.414.2716B}. 
This modified version of \textsc{swift} has been made available to the community.\footnote{\url{https://sirrah.troja.mff.cuni.cz/yarko-site/tmp/dt/swift_rmvsy.html}} 
On the other hand, several different effects have been added in the \textsc{rebound}\footnote{\url{https://rebound.readthedocs.io/en/latest/}} integrator \citep{2020MNRAS.491.2885T}, but the Yarkovsky and YORP effects are still not among those.


In this paper, we describe our implementation of the Yarkovsky/YORP effect in the \textsc{mercury} and \textsc{orbit9} integrators. Both integrators have been extensively used for the simulations of the long-term orbit evolution of asteroids. For recent examples of \textsc{mercury} based simulations see e.g. \citet{2020AJ....159..179H,2020MNRAS.499.4626D,2021MNRAS.506L...6M}, while for recent works based on \textsc{orbit9} see e.g. \citet{2020MNRAS.497.4921K,2021MNRAS.501..356P,2021MNRAS.505.1917D}.
Despite being extensively used, the Yarkovsky and YORP effects are not included in the \textsc{mercury} package. The \textsc{orbit9} integrator does not include the YORP effect, while the Yarkovsky was implemented using a very simplified formulation \citep[see e.g.][]{2015ApJ...807L...5N}.


To implement the two effects in the \textsc{mercury} and \textsc{orbit9} integrators we adopted the following strategy. The Yarkovsky effect is computed as a non-conservative vector field along the tangential direction, that produces the semi-major axis drift given by analytical theories \citep[see, e.g.][]{vokrouhlicky_1999}. For the YORP effect, we follow a similar approach to the one described in \citet{2015Icar..247..191B}, which is suitable for statistical studies of the evolution of asteroids. In our code, we provide two different YORP models: a static and a stochastic one.  Along with the two effects, we introduced the effect of collisions and breakups on the asteroid spin state.
%

Finally, in addition to the details on the implementation, we performed tests to validate the new codes. In this respect, we first show that the integrations obtained with the two packages agree with each other. We also provide some additional tests to show how the static and stochastic YORP model works.

The extended version of the \textsc{mercury} integrator is publicly available at \url{https://github.com/Fenu24/mercury}, while the modified version of \textsc{orbit9} can be found at \url{https://github.com/Fenu24/OrbFit}.

\section{THE \textnormal{\textsc{mercury}} AND \textnormal{\textsc{orbit9}} INTEGRATORS}
In this section we provide more details about \textsc{mercury} and \textsc{orbit9} $N$-body software packages.

\subsection{The \textnormal{\textsc{mercury}} package}
The \textsc{mercury}\footnote{\url{https://www.astro.keele.ac.uk/~dra/mercury/}} 
integrator package is written in \texttt{fortran} language, and it contains several integration schemes: a general Bulirsch-Stoer algorithm \citep[see e.g.][]{bulirsch-stoer_2002}, a Radau integrator \citep{everhart_1985}, a second order mixed-variable symplectic (MVS) integrator \citep{1996FIC....10..217W}, and a hybrid symplectic integrator \citep{1999MNRAS.304..793C} that is able to handle close encounters with planets.

\textsc{mercury} is designed to compute the evolution of objects moving around a massive central body, such as planets and asteroids around the Sun, a system of moons around a planet, or exo-planetary systems. The integrators include the gravitational forces due to the central mass, and some other specified massive bodies. It can also compute non-gravitational forces for comets \citep{1973AJ.....78..211M}, or compute the effects of the gravitational $J_2, J_4, J_6$ moments of the central body. 

Other forces depending on the positions and velocities can be added by editing the subroutine \texttt{mfo\_user}. The Yarkovsky effect is directly implemented here as described in Sec.~\ref{s:yarkovsky}. 
The integration of the spin-axis dynamics is performed in the main integration loop of the
orbital dynamics. The spin-axis and the rotation rate evolve on a time scale much larger
than the orbital period, hence a longer time-step is used. At each integration step of the
orbital dynamics, we check whether an integration step for the spin-axis dynamics is needed or not. In the positive case, the obliquity and the rotation rate are updated as described in Sec.~\ref{s:yorp}, and the semi-major axis drift due to the Yarkovsky effect is recomputed. The subroutines used for these tasks are all implemented in an external module named \texttt{yorp\_module}. 

On a more technical side, \textsc{mercury} is capable to detect collisions between a
planet (or the central body) and a massless object, or to detect an escape from the
system. When either of these occur, the array containing the IDs of objects is re-sized
and re-arranged in order to remove the colliding (or escaping) particle from the
integration. The indexing of the objects is therefore changed at every removal event. 
The arrays containing all the parameters needed for the coupled Yarkovsky/YORP effect
dynamics are re-arranged as well, according to the new objects' indexing.


\subsection{The \textnormal{\textsc{orbit9}} integrator from the \textnormal{\textsc{orbfit}} package}
The \textsc{orbit9} integrator is included in the \textsc{orbfit}\footnote{\url{http://adams.dm.unipi.it/orbfit/}} package, and it is also written in \texttt{fortran} language. It includes three different integration schemes: a Radau integrator \citep{everhart_1985}, a symplectic Runge-Kutta-Gauss with variable orders \citep[see e.g.][]{hairerII}, and a multistep method with symplectic starter \citep{1988CeMec..43....1M}.

This package is explicitly designed for the long-term integration of main belt asteroids, and it includes the gravitational attraction of the planets. In the case only the giant gas planets are included in the model, a barycentric correction can be applied to take into account the effects of the missing inner planets.
The integrator permits to add to the dynamical model also the quadrupole $J_2$ moment of the Sun, and relativistic effects. Details of the implementation can be found in \citet{1989A&A...210..313N}.
This package also includes several useful tools, such as an on-line removal of short periodic effects \citep[][]{1987A&A...181..182C}, close approaches monitoring, and the computation of the maximum Lyapounov Characteristics Exponents. Due to all these options, \textsc{orbit9} is widely used for the computation of proper elements of main-belt asteroids \citep{2000CeMDA..78...17K}. 

Additional forces can be easily added in the subroutine \texttt{force9}. \textsc{orbit9} already provides a simple implementation of the Yarkovsky effect, in which the semi-major axis drift is provided in input for each asteroid, and then kept constant over the whole integration time-span. For the purpose of this work, we replaced this part with our implementation described in Sec.~\ref{s:yarkovsky}. To include the YORP effect, we used a similar solution to the one described for \textsc{mercury}, and the subroutines for this purpose are all contained in a module called \texttt{yorp\_module}.

\section{YARKOVSKY EFFECT IMPLEMENTATION}
\label{s:yarkovsky}
The Yarkovsky effect mainly causes a change in semi-major axis, and the drift
   $\textrm{d}a/\textrm{d}t$ depends on several orbital and physical parameters of the
   asteroid, i.e.
\begin{equation}
    \frac{\textrm{d}a}{\textrm{d}t} = \bigg(\frac{\textrm{d}a}{\textrm{d}t}\bigg)(a,D,\rho, K,C,\gamma, P, \alpha, \varepsilon).
    \label{eq:drift_dep}
\end{equation}
In Eq.~\eqref{eq:drift_dep}, $a$ is the semi-major axis of the orbit, $D$ is the diameter
of the asteroid, $\rho$ is the density, $K$ is the thermal conductivity, $C$ is the heat
capacity, $\gamma$ is the obliquity, $P$ is the rotation period, $\alpha$ is the
absorption coefficient, and $\varepsilon$ is the emissivity. The total semi-major axis
drift is given by the sum of the diurnal and the seasonal effects, i.e.
\begin{equation}
    \frac{\textrm{d}a}{\textrm{d}t} = \kappa_1 \cos \gamma + \kappa_2 \sin^2\gamma,
    \label{eq:yarkoDrift}
\end{equation}
where $\kappa_1$ and $\kappa_2$ are functions of the asteroid orbit, physical parameters,
and rotation period \citep[see e.g.][Appendix A, for details]{1998A&A...338..353V,
vokrouhlicky_1999, 2021A&A...647A..61F}.
The acceleration due to the gravitational attraction of the planets is then augmented with a force of the form $F = F_\tau \tau$, where $\tau$ is the tangential direction. The multiplying factor $F_\tau$ is obtained through the Gauss planetary equations \citep[see e.g.][]{murray_and_dermott} as
\begin{equation}
    F_\tau = \frac{\textrm{d}a}{\textrm{d}t}  \frac{\sqrt{GM_\odot (1-e^2)}}{2 a^{3/2}(1+e\cos f)},
    \label{eq:yarkoMag}
\end{equation}
where $G$ is the universal gravitational constant, $M_\odot$ is the mass of the Sun, $e$ is the orbital eccentricity, and $f$ is the true anomaly. 

The Gauss equation for $\text{d}a/\text{d}t$ also contains the contribution of the radial
component $F_r$, multiplied by a factor $e\sin f$. The eccentricity of main-belt asteroids
is generally small; therefore, we neglected the term containing $F_r$ in the derivation of
Eq.~\eqref{eq:yarkoMag} since its contribution is usually much smaller than that of
$F_\tau$.
Note that the model of the Yarkovsky drift of Eq.~\eqref{eq:yarkoDrift} is
obtained by assuming a circular orbit of the asteroid. While it has been proved that the
magnitude of the Yarkovsky effect changes even by an order of magnitude when $e$ passes
from $\sim$0.1 to $\sim$0.9 \citep{2001Icar..149..222S}, eccentricities up to $\sim$0.3
produce variations only up to $\sim$20\% percent with respect to the analytical circular
model.

For each small object integrated, the values of $D$, $\rho$, $K$, $C$, $\alpha$, and
$\varepsilon$ are provided in input, and they are kept constant throughout the whole
integration time-span. We also provide in input the initial values of the obliquity
$\gamma$, and of the rotation period $P$, that are used to compute the initial semi-major
axis drift, and to set the initial conditions for the spin-axis dynamics.

\section{YORP EFFECT IMPLEMENTATION}
\label{s:yorp}
The strength of the semi-major axis drift of Eq.~\eqref{eq:yarkoDrift} due to the
Yarkovsky effect depends on both the obliquity $\gamma$ and the rotation period $P$. In
turn, thermal effects produce a non-zero torque known as YORP effect \citep[see
e.g.][]{bottke-etal_2006}, that is able to change both $\gamma$ and $P$. The YORP effect
sensitively depends on small surface features, and a precise determination can be done
only when a good shape model is known \citep[][]{vokrouhlicky-etal_2015}. On the other
hand, a Monte Carlo approach can be used to determine the torques for the purpose of
statistical studies, as we consider in this work. 
The evolution of the obliquity $\gamma$ and the rotation rate $\omega = 1/P$ are defined
by
\begin{equation}
    \begin{cases}
    \displaystyle\frac{\textrm{d}\omega}{\textrm{d}t} & = f(\gamma), \\[2ex]
    \displaystyle\frac{\textrm{d}\gamma}{\textrm{d}t} & = \displaystyle\frac{g(\gamma)}{\omega},
    \end{cases}
    \label{eq:YORPvf}
\end{equation}
where $f,g$ are the torques. The selection of these functions is discussed in
Sec.~\ref{ss:choose_fg}.
Eq.~\eqref{eq:YORPvf} is integrated numerically with an explicit Runge-Kutta method of
order 4 \citep[see e.g.][]{bulirsch-stoer_2002}, using a constant time step $\Delta
t_{\text{spin}}$. The time step can be either specified by the user or set automatically.
The automatic selection is implemented as a simple re-scaling with the minimum diameter
$D_{\min}$ among the small bodies, i.e.
\begin{equation}
    \Delta t_{\text{spin}} = 50 \text{ yr} \times \frac{D_{\min}}{1 \text{ km}}. 
    \label{eq:deltaT_spin}
\end{equation}
To avoid too short or too long time steps, we set a lower limit of 1 yr and an upper limit
of 50 yr. 
Note that the magnitude of the YORP effect scales as $1/D^2$ (see
Eq.~\ref{eq:YORPscaleFactor}), still the median YORP timescale for
objects of $\sim$50 m in diameter is of the order of few thousands years
\citep[see e.g.][]{2004Icar..172..526C,2021AJ....162..227F}. The re-scaling of
Eq.~\eqref{eq:deltaT_spin} is chosen to slightly reduce the time step for 
asteroids smaller than 1 km in diameter, while maintaining a good compromise between the
accuracy and the speed of the numerical integration.

The YORP effect couples the spin dynamics of the asteroid with the orbital dynamics
through the Yarkovsky effect. For this reason, at each integration step of
Eq.~\eqref{eq:YORPvf} we also update the value of the semi-major axis drift
$\textrm{d}a/\textrm{d}t$ of Eq.~\eqref{eq:yarkoDrift}, by using the new values of
$\gamma$ and $P$ obtained from the spin-axis dynamics.

An important issue to deal with when modeling the spin-axis evolution by the YORP effect
is handling the asymptotic states. At the asymptotic states, the rotation period becomes
either very long or very short, and the currently available YORP effect models are not
valid for such extreme cases. The process of an asteroid entering and eventually exiting
from an asymptotic state with a new spin state is generally called YORP cycle. We discuss
how to handle YORP cycles in Sec.~\ref{ss:long_period} and ~\ref{ss:short_period}.

\subsection{Choosing the torques}
\label{ss:choose_fg}
Accurate shape models are known only for a small number of asteroids, and therefore they cannot be used for the purpose a statistical characterization of the YORP torques.
\cite{1998A&A...332.1087M} introduced the concept of Gaussian random spheres, that are suitable for generating synthetic asteroid shapes. Using this representation, \cite{2002Icar..159..449V} computed the thermal torques on a  sample of 500 asteroids with mean diameter $D_0=2$ km, bulk density $\rho_0 = 2500$ kg m$^{-3}$, and placed on a heliocentric circular orbit with radius $a_0 = 2.5$ au, using the approximation of zero thermal conductivity. Later on, \cite{2004Icar..172..526C} added the effect of non-zero thermal conductivity in the YORP effect model, and computed the torques for a sample of 200 objects in the cases of conductivity $0.001$ W m$^{-1}$ K$^{-1}$ and $0.01$ W m$^{-1}$ K$^{-1}$. Generally, the obliquity $\gamma$ reaches an extreme value of 0, 90, or 180 deg, while the rotation is either accelerated or decelerated.

The statistics for the asymptotic states is sensible to the chosen conductivity. In the case of zero thermal conductivity, \cite{2002Icar..159..449V} found that there is approximately the same probability for $\gamma$ to reach 0/180 deg or 90 deg, while the rotation is slowed down asymptotically for the overwhelming majority of the objects.
For $K = 0.001$ W m$^{-1}$ K$^{-1}$ \citet{2004Icar..172..526C} found that about 80\% of the objects are driven towards 0/180 deg in obliquity, while 40\% of them asymptotically accelerate the rotation rate. 
Finally, the authors found that for $K = 0.01$ W m$^{-1}$ K$^{-1}$ about 95\% of the objects reach 0/180 deg in obliquity, and there is about the same probability to accelerate or decelerate the rotation rate. 

\begin{figure}[!ht]
    \centering
    \includegraphics[width=0.48\textwidth]{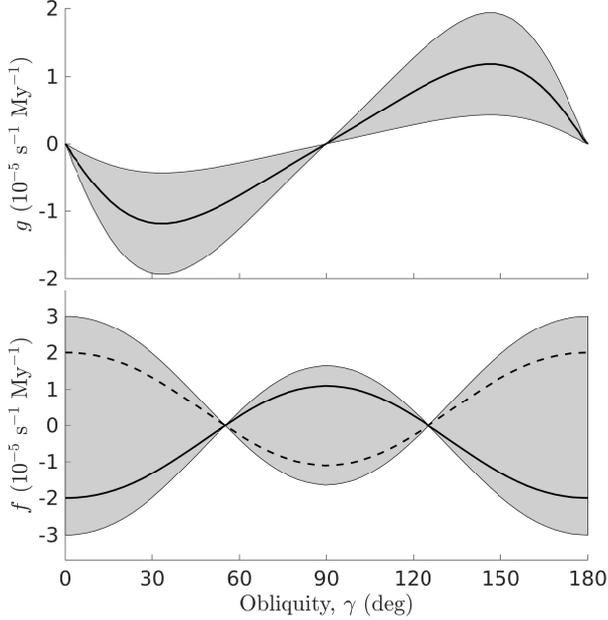}
    \caption{Torques $f,g$ as function of the obliquity $\gamma$ by
       \citet{2004Icar..172..526C}, in the case $K=0.01$ W m$^{-1}$ K$^{-1}$. The black
       solid curves are the average torques computed over a set of 200 Gaussian random spheres of diameter $D_0 = 2$ km, density $\rho_0 = 2500$ kg m$^{-3}$, and placed on circular heliocentric orbits with radius $a_0 = 2.5$ au. The gray zone defines the variance of the results.}
    \label{fig:fg_fun}
\end{figure}

Figure~\ref{fig:fg_fun} shows the average torques $f,g$ and their variance, computed by \citet{2004Icar..172..526C} in the case $K = 0.01$ W m$^{-1}$ K$^{-1}$. Note that $g$ is anti-symmetric with respect to 90 deg, and that it is negative in the interval $(0, 90)$ deg. This means that, in this case, the asymptotic values for the obliquity are 0 or 180 deg. On the other hand, $f$ is symmetric with respect to 90 deg and the values at 0 and 180 deg can be either positive or negative, meaning that the objects may accelerate or decelerate asymptotically.

In our implementation, the choice of the functions $f,g$ is drawn from the average and variance values shown in Fig.~\ref{fig:fg_fun}, and from the statistics presented in \citet{2004Icar..172..526C}. Since the computations were not performed on a large enough set of conductivity values, we distinguish two cases: if $K \leq 0.005$ W m$^{-1}$ K$^{-1}$ we use the distribution of torques obtained for $K = 0.001$ W m$^{-1}$ K$^{-1}$, while we use the distribution obtained for $K = 0.01$ W m$^{-1}$ K$^{-1}$ otherwise. 

For the case $K = 0.001$ W m$^{-1}$ K$^{-1}$, we randomly select a function $g$ from the gray area in Fig.~\ref{fig:fg_fun}, and then we decide whether to change its sign or not, according to the 80\% probability of reaching 0/180 deg in obliquity. On the other hand, a function $f$ from the gray area is randomly chosen, where the asymptotically accelerating case is selected with 40\% probability.

For $K = 0.01$ W m$^{-1}$ K$^{-1}$, we make the simplified assumption that all the objects reach 0/180 deg in obliquity, and that half of the objects accelerate the rotation rate. Hence, here we simply randomly choose a pair $f,g$ from the corresponding gray areas, without any further constraints.  

Once $f,g$ are chosen, they also need to be re-scaled to account for different diameter $D$, bulk density $\rho$, and semi-major axis $a$. According to \citet{2011MNRAS.414.2716B}, the re-scaling factor is given by 
\begin{equation}
    c = \bigg( \frac{a_0}{a} \bigg)^2 \bigg( \frac{D_0}{D} \bigg)^2 \bigg( \frac{\rho_0}{\rho} \bigg)c_{\text{YORP}}.
    \label{eq:YORPscaleFactor}    
\end{equation}
In Eq.~\eqref{eq:YORPscaleFactor}, $c_{\text{YORP}}$ is a parameter that accounts for uncertainties in the YORP modeling. \citet{2015Icar..247..191B} estimated this parameter by calibrating the model with the available observations, and constrained it to be in the range 0.5-0.7. This means that the YORP effect is actually less efficient than what predicted by the model. In our code, $c_{\text{YORP}}$ is included as an input parameter that can be chosen by the user, while the default value is set to $c_{\text{YORP}} = 0.7$.

We stress again that the torques $f$ and $g$ of Fig.~\ref{fig:fg_fun} were obtained by
assuming a circular orbit for the asteroid. \citet{2007Icar..188..430S} found that the
YORP torques averaged over an orbital period scale with the eccentricity $e$ with a factor
proportional to $1/\sqrt{1-e^2}$, that behaves as $1+\mathcal{O}(e^2)$ for nearly circular
orbits. Thus, the effects of a small non-zero eccentricity result to be negligible, and
the YORP model we adopted in the code should be suitable for main-belt asteroids.

\subsection{Long rotation period}
\label{ss:long_period}
When an asteroid rotates very slowly, it could enter in a tumbling rotation state
\citep[see e.g.][]{2007Icar..191..636V}. In this case, the Yarkovsky/YORP effect models by
\citet{1998A&A...338..353V, 2004Icar..172..526C} are not valid anymore, because they
assume the body to rotate around a principal axis. 
In our implementation, the spin-axis evolution is stopped when the rotation period is such
that $P > 1000$ h, and the last recorded state is kept constant as time increases. Note
that this has a small significance for the Yarkovsky effect, since the magnitude of the
semi-major axis drift is minimized for very long rotation period.  

In this extreme case, we assume that the spin state is re-initialized by a
sub-catastrophic collision capable to re-orient the body. The characteristic timescale
$\tau_{\text{reor}}$ for such collision is \citep[see eg.][]{1998Icar..132..378F,
2006Icar..182..118V}
\begin{equation}
    \tau_{\text{reor}} =  B \bigg( \frac{\omega}{\omega_0} \bigg)^{\beta_1}\bigg( \frac{D}{D_0} \bigg)^{\beta_2}c_{\text{reor}},
    \label{eq:tau_reor}
\end{equation}
where $B = 84.5$ kyr, $\omega_0 = 1/5$ h$^{-1}$, $D_0 = 2$ m, $\beta_1 = 5/6$, and
$\beta_2 = 4/3$. The parameter $c_{\text{reor}}$ accounts for the uncertainties in the
timescale estimation. \citet{2015Icar..247..191B} constrained this parameter by
calibrating the YORP model with available observations, and found $c_{\text{reor}}=0.9$.
In the code, this parameter can be specified in input by the users, and we suggest to set
it to the nominal estimated value of 0.9.
The collision re-orientation is modeled as an
uncorrelated and random Poissonian process with parameter $\tau_{\text{reor}}$. More
specifically, we compute $\tau_{\text{reor}}$ using Eq.~\eqref{eq:tau_reor} at each
integration timestep of the spin-axis dynamics. Then, we generate a random number $n$
between 0 and 1, and we assume that a collision occurs if $n < 1-\exp{(-\Delta
t_{\text{spin}}/\tau_{\text{reor}})}$. 
%


When a sub-catastrophic collision takes place, $\gamma$ is randomly re-initialized
according to a random orientation in space, i.e. $\cos\gamma$ uniformly distributed
between $-1$ and $1$. The rotation period $P$ is also re-initialized, however choosing its
distribution is not straightforward. Indeed, \citet{2002aste.book..113P} found that
the spin-state of asteroids larger than 40 km in diameter is compatible with a Maxwellian
distribution with peak at 8 h, however this property may change at smaller diameters \citep[see
e.g.][]{1998Icar..132..378F}. We therefore use a Maxwellian distribution to
re-initialize $P$, leaving the value of the peak as a free parameter to be chosen by the
user.

In addition, \citet{2009Icar..202..502S} evaluated how the YORP effect changes when a crater is added
to the surface of an asteroid. The author shown that a crater as large as 0.6 times the
mean radius of the asteroid could produce an error of 100\% in the torques, with respect
to the torques computed without the crater.
For this reason, we choose new functions $f,g$ for the spin-axis evolution at each
sub-catastrophic re-orientation event, as described in Sec.~\ref{ss:choose_fg}.

\subsection{Short rotation period}
\label{ss:short_period}
\begin{figure}[!ht]
    \centering
    \includegraphics[width=0.48\textwidth]{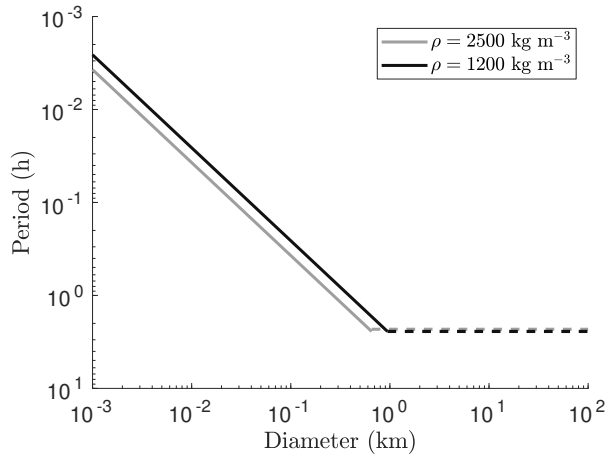}
    \caption{Limiting rotation period for two different densities computed using a bulk cohesion of $100$ Pa.}
    \label{fig:limiting_period}
\end{figure}

Asteroids rotating very fast may change shape, shed mass, or undergo fission to form a
binary system. Objects smaller than about 150 meters in diameter could rotate very fast
\citep{2000Icar..148...12P}, and the critical spin limit depends on both the size and the
internal cohesive forces \citep[see e.g.][]{2007Icar..187..500H, 2021MNRAS.502.5277H}. In
this work, we use the critical rotation period given by \citet{2021MNRAS.502.5277H} for
small size, corresponding to
\begin{equation}
    P_{\text{crit}} = 2\pi\sqrt{\frac{\rho k}{5C}}D.
    \label{eq:criticalPeriod}
\end{equation}
In Eq.~\eqref{eq:criticalPeriod}, $k = 0.9114$ is a parameter computed with friction angle equal to $32.5$ deg, and $C$ is the bulk cohesion. 
Because Eq.~\eqref{eq:criticalPeriod} is not valid at large size, we set a hard limit of
$P_{\text{crit}} = 2.44$ h when Eq.~\eqref{eq:criticalPeriod} gives too long periods. This
value corresponds to the critical period found by \citet{2000Icar..148...12P}. 
Selecting an appropriate bulk cohesion for Eq.~\eqref{eq:criticalPeriod} is not a simple
task, due to the lack of observational constraints. \citet{sanchez-scheeres_2018}
estimated the bulk cohesion of rubble-pile asteroids Ryugu, 1950 DA, and 2008 TC3 to be in
the range 10-100 Pa. More recently, \citet{2021Icar..36214433Z} estimated the minimum bulk
cohesion of Dydimos to be of the order of 10 Pa. On the other hand, values of the order of
kPa and larger are expected to be typical for monolithic bodies.
In our code, we adopted a constant bulk cohesion $C=100$ Pa, which could be an appropriate
value for rubble-piles.
Figure~\ref{fig:limiting_period} shows the critical period for $\rho = 1200$ kg m$^{-3}$
and $\rho = 2500$ kg m$^{-3}$. Note that this is a fairly good approximation of the
limiting period found by \citet{2007Icar..187..500H} using the continuum theory.

When the rotation period becomes smaller than $P_{\text{crit}}$ we assume that a fission
event takes place, and we re-initialize the spin state. In this process, we do not
simulate the production of a binary system, and we assume that the mass lost is small
enough to not significantly change the equivalent diameter of the object. During the
fission event, we assume the obliquity $\gamma$ to be unchanged. On the other hand, the
spin rate is decreased according to the momentum carried away by the ejected mass
\citep[see][]{2010Natur.466.1085P}, and it is given by
\begin{equation}
    \omega^2_{\text{new}} = \omega^2_{\text{old}} - kq.
    \label{eq:newspin}    
\end{equation}
In Eq.~\eqref{eq:newspin}, $k = (20\pi G/6) \times (2000 \text{ kg m}^{-3})$ is a constant
\citep[see the Supplementary Material of][]{2010Natur.466.1085P}, and $q$ is the mass
ratio of the ejected mass to the asteroid mass. The mass ratio $q$ is chosen randomly at
each fission event in the range 0.002-0.2, in such a way that $\log q$ is uniformly
distributed. 

\citet{2009Icar..202..502S} evaluated how the YORP effect changes when adding boulders on
the surface of an asteroid. The author indicated that the torques are affected by the
distribution of both small and large boulders, and that a single large boulder can already
seriously change the YORP effect. During a spin-up event, boulders can experience
landslide or fission, and these phenomena all contribute in changing the torques. 
For these reasons, during a fission event we also select new functions $f,g$ as described
in Sec.~\ref{ss:choose_fg}. 

It is important to remark that small-scale surface features of sub-kilometer rubble-pile
asteroids are largely dominated by boulders. This was already discovered by the JAXA
Hayabusa mission on asteroid Itokawa \citep{2008EP&S...60...13M}, and it has been
confirmed more recently by the NASA OSIRIS-REx and the JAXA Hayabusa2 missions on the
carbonaceous asteroids Bennu and Ryugu \citep{2019Natur.568...55L, 2019Sci...364..268W}.
Moreover, the presence of boulders seems to be independent from the spectral type. It is
not clear yet whether these surface features are preserved at sizes smaller than about 150
meters in diameter or not, especially when the rotation is very fast. Theoretical models
by \citet{2020Icar..33813443S} suggest that gravel and boulders can survive on the surface
of small monolithic fast rotators. Additionally, estimations of thermal inertia on the
small super-fast rotator (499998) 2011 PT \citep{2021A&A...647A..61F} suggest either a
regolith covering or a rubble-pile structure. These works indicate that the re-shaping
assumption at fast rotation may be a good choice even at diameters smaller than 150
meters. 

\section{STOCHASTIC YORP EFFECT}
\label{s:stoc_yorp}
The YORP effect model of Sec.~\ref{s:yorp} is also referred to as a static YORP model by \citet{2015Icar..247..191B}. The analysis on the sensitivity of the YORP effect on small-scale surface features performed by \citet{2009Icar..202..502S} suggested that the spin evolution could be stochastic rather than deterministic, and this hypothesis has been supported later by hard-sphere discrete element numerical simulations by \citet{2015ApJ...803...25C}. 
A simpler stochastic YORP model has also been introduced by \citet{2015Icar..247..191B}. 
Thermal torques dramatically depend on the surface distribution of both boulders and craters. However, these two distributions are not steady with time. Boulders can migrate by the effect of spinning-up, while their size distribution is changed by cracking induced by micro-meteoroid impacts \citep{horz-cintala_1997}, and by thermal effects \citep{delbo-etal_2014}. During the artificial impact experiment performed by the Hayabusa2 mission on asteroid Ryugu, a 2 kg copper projectile was shot at 2 km s$^{-1}$. It created a $\sim$15 m class crater \citep{2020Sci...368...67A}, demonstrating that impacts could significantly contribute to form craters on rubble-piles and to move boulders across the surface. Additionally, close encounters with planets could also modify the overall shape of the body by the effect of tides \citep{1998Icar..134...47R, 2021Icar..35814205K}.
These small and large-scale alterations in shape contribute to a stochastic change of the
spin-axis evolution onto a new set of YORP curves.
Therefore, the assumption that the thermal torques remain constant during an entire YORP cycle made in Sec.~\ref{s:yorp} is not valid anymore, but on the contrary, they should change with a shorter timescale.

To implement the stochastic YORP model, we use an approach similar to \citet{2015Icar..247..191B}. A size-dependant timescale $\tau_{\text{YORP}}(D)$ is introduced, and the torques $f,g$ are changed after a time $\tau_{\text{YORP}}(D)$ has passed, without modifying the spin state. Using estimates on the characteristic timescale of cratering events, \citet{2015Icar..247..191B} found a value of $\tau_{\text{YORP}} = 1$ My for asteroids with size $D \sim 4$-$8$ km. In the same work, the authors adopted $\tau_{\text{YORP}} = 0.5$ My for asteroid (175706) 1996 FG3 ($D = 1.9$ km), and $\tau_{\text{YORP}} = 0.25$ My for asteroid Ryugu ($D = 0.89$ km). 
Since the collision re-orientation timescale of Eq.~\eqref{eq:tau_reor} depends on the diameter with an exponent of $\beta_2 = 4/3$, and these collisions generally change the shape, we assume for $\tau_{\text{YORP}}$ the same size dependency. Therefore, in our code we use 
\begin{equation}
    \tau_{\text{YORP}}(D) = 0.25\text{ My} \times \bigg(\frac{D}{1\text{ km}} \bigg)^{4/3}\times c_{\text{stoc}}, 
    \label{eq:tau_yorp}
\end{equation}
where $c_{\text{stoc}}$ is a parameter that can be used to take into account the uncertainties on this timescale, and it can be specified in input. To roughly match the values used by \citet{2015Icar..247..191B}, we suggest to set this parameter to $c_{\text{stoc}} = 0.75$.

At size of $\sim$1 km, this timescale is several factor smaller than the YORP cycle timescale, so that 6-10 re-shaping events are expected during a whole cycle.  
As the size decreases, the YORP cycle timescale scales as $D^2$, while $\tau_{\text{YORP}}$ of Eq.~\eqref{eq:tau_yorp} scales as $D^{4/3}$, meaning that there may be a limiting diameter below which the stochastic YORP does not introduce drastic changes in the spin-axis evolution. 
However, it is not clear yet whether the trend in $\tau_{\text{YORP}}$ is valid down to sizes of $~$10 m in diameter or not, and this requires more investigation in the future.

In the higher conductivity case $K = 0.01$ W m$^{-1}$ K$^{-1}$, the stochastic YORP does not modify the asymptotic obliquity states, and objects are driven towards 0 or 180 deg. On the other hand, the sign of $f$ at 0/180 deg randomly changes at each stochastic event, meaning that the rotation could either slow down or speed up. This has the effect that the rotation rate $\omega$ evolves as a random-walk process, which generally increases the timescale of the YORP cycle.
In the lower case $K = 0.001$ W m$^{-1}$ K$^{-1}$, there is a 20\% probability to choose function $g$ with 90 deg as asymptotic obliquity, that is reached by slowing down the rotation. Therefore, here the spin-axis evolution could be even more complicated, since also the obliquity $\gamma$ may behave as a random-walk process as well.

\section{TESTS}
In this section, we provide some tests of the developed codes. First, we show that the
integrations performed with \textsc{mercury} and \textsc{orbit9} produce results that are
in excellent agreement. Later, we provide some tests to show how the static and stochastic
YORP models behave. 

\subsection{\textnormal{\textsc{mercury}} vs. \textnormal{\textsc{orbit9}}: single object comparison}
\label{ss:test1}
\begin{figure*}[!ht]
    \centering
    \includegraphics[width=0.9\textwidth]{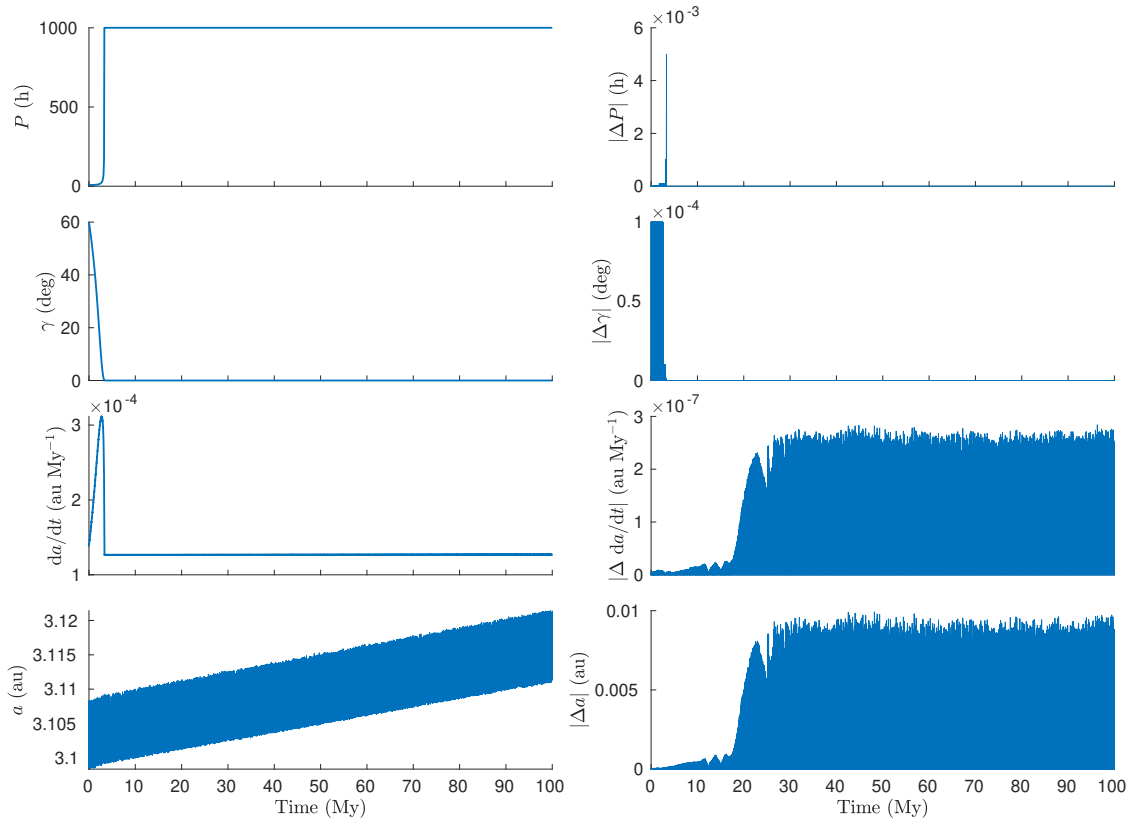}
    \caption{Parameters evolution of the test particle with diameter $D = 2$ km, for the
       conductivity case $K =  0.01$ W m$^{-1}$ K$^{-1}$. The left column shows, from the
       top to the bottom, the evolution of the rotation period $P$, the obliquity
       $\gamma$, the semi-major axis drift $\text{d}a/\text{d}t$, and the semi-major axis
       $a$. The right column shows the evolution of the absolute difference between these
       quantities obtained with \textsc{mercury} and \textsc{orbit9}.}
    \label{fig:comparison_mercury_orbit9}
\end{figure*}

We first perform a test to verify that the coupled Yarkovsky/YORP effect implementations
of \textsc{mercury} and \textsc{orbit9} give consistent results. We integrated the
dynamics of 6 test particles for 100 My, taking into account only the gravitational
attraction of Jupiter, Saturn, Uranus, and Neptune. 
The initial orbit of the particles has a semi-major axis of 3.1 au, a low eccentricity of
0.01, and an inclination of 1 deg, so that possible chaotic effects are limited.  
The sizes were set to $D = 2, 1$ and $0.5$ km, and we used two different values of thermal
conductivity, namely $K = 0.001, 0.01$ W m$^{-1}$ K$^{-1}$.
We assumed an initial obliquity of $60$ deg and an initial rotation period of $8$ h for
all the objects.
Moreover, we used a density of $1200$ kg m$^{-3}$, heat capacity of $800$ J m$^{-1}$
K$^{-1}$, and values of absorption coefficient and emissivity equal to $1$. 
We chose the MVS integration method for the simulations performed with \textsc{mercury},
and the multi-step integration method for the simulations carried on with \textsc{orbit9}.
For both of them, we used a constant time step of 5 days for the orbital evolution and a
constant time step $\Delta t_{\text{spin}} = 50$ yr for the spin-axis evolution. 

To make a coherent comparison between the evolution obtained with the two different
packages, we forced the torques $f$ and $g$ to be equal to the mean torques (see
Fig.~\ref{fig:fg_fun}, solid black line).  We also switched off all the re-orientations
and re-shaping events described in Sec.~\ref{s:yorp} and \ref{s:stoc_yorp}. In this
manner, the spin-axis stops evolving when an end-state is reached.

The left column of Fig.~\ref{fig:comparison_mercury_orbit9} shows the time evolution of
the rotation period $P$, the obliquity $\gamma$, the semi-major axis drift
$\text{d}a/\text{d}t$, and the semi-major axis $a$, for the asteroid with $D = 2$ km and
$K =  0.01$ W m$^{-1}$ K$^{-1}$. The rotation of this object is slowed down until it
reaches a period of 1000 h, while the obliquity evolves towards 0 deg. The asymptotic
state is reached after about 3.5 My of evolution. The semi-major axis drift increases during
the first period of evolution by the effect of the obliquity moving towards an extreme
value. Then it suddenly drops by the effect of the increasing rotation period. In the
bottom panel, we can appreciate the increasing trend of the semi-major axis $a$.

The right column of Fig.~\ref{fig:comparison_mercury_orbit9} shows the absolute
differences $|\Delta P|, |\Delta \gamma|, |\Delta \text{d}a/\text{d}t|, |\Delta a|$
between the evolution computed with \textsc{mercury} and the one computed with
\textsc{orbit9}. From these plots, we can appreciate that the results obtained with the
two packages are in good agreement. Indeed, the maximum difference in the rotation period
   is of the order of $10^{-3}$ h, and $10^{-4}$ deg in obliquity. The difference in the
semi-major axis is smaller than $10^{-3}$ au during the first $\sim$17 My evolution,
while it grows larger after 20 My, though never exceeding 0.01 au. However, these 
differences after 20 My may be caused either by long-term integration errors, by the fact
that the integration method for the orbital evolution is not the same, or by some
small chaotic and diffusion effects.
Differences in the rotational dynamics do not play a role here because the spin-axis has
already reached the end-state at this point of evolution. 
Finally, the difference in the semi-major axis drift is
of the order of $10^{-7}$ au My$^{-1}$, which is orders of magnitude smaller than the
typical drifts of km-sized asteroids and smaller. 

Similar results were obtained for the other 5 test particles and therefore are not
reported here. These tests show that the two implementations produce compatible results,
and therefore they can be considered equivalent.

\subsection{\textnormal{\textsc{mercury}} vs. \textnormal{\textsc{orbit9}}:  static and stochastic YORP models}
\label{ss:test2}
\begin{figure*}[!ht]
    \centering
    \includegraphics[width=\textwidth]{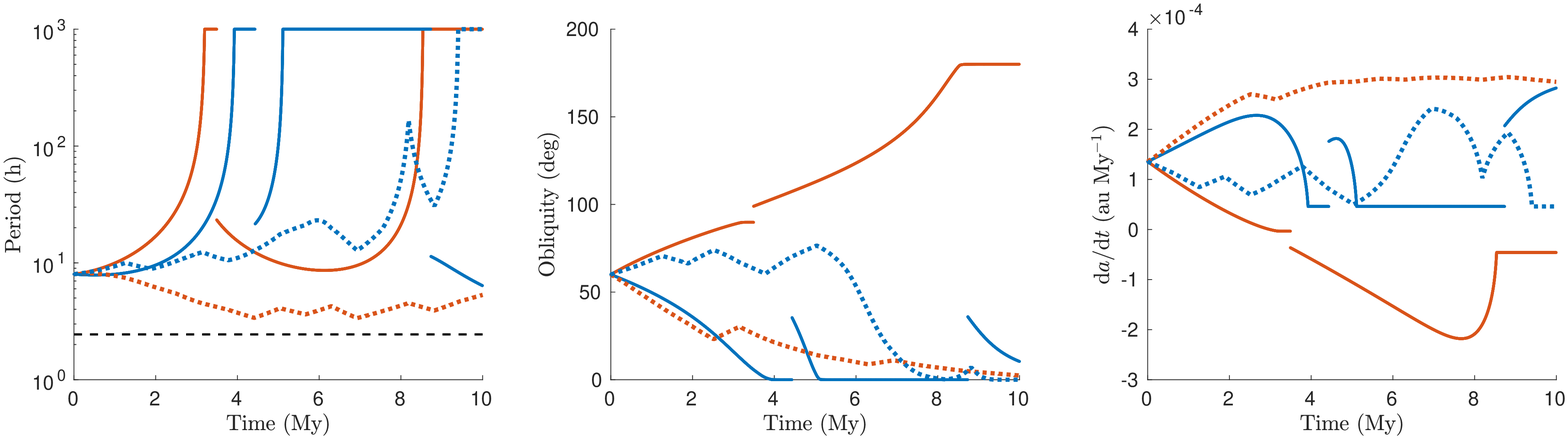}
    \includegraphics[width=\textwidth]{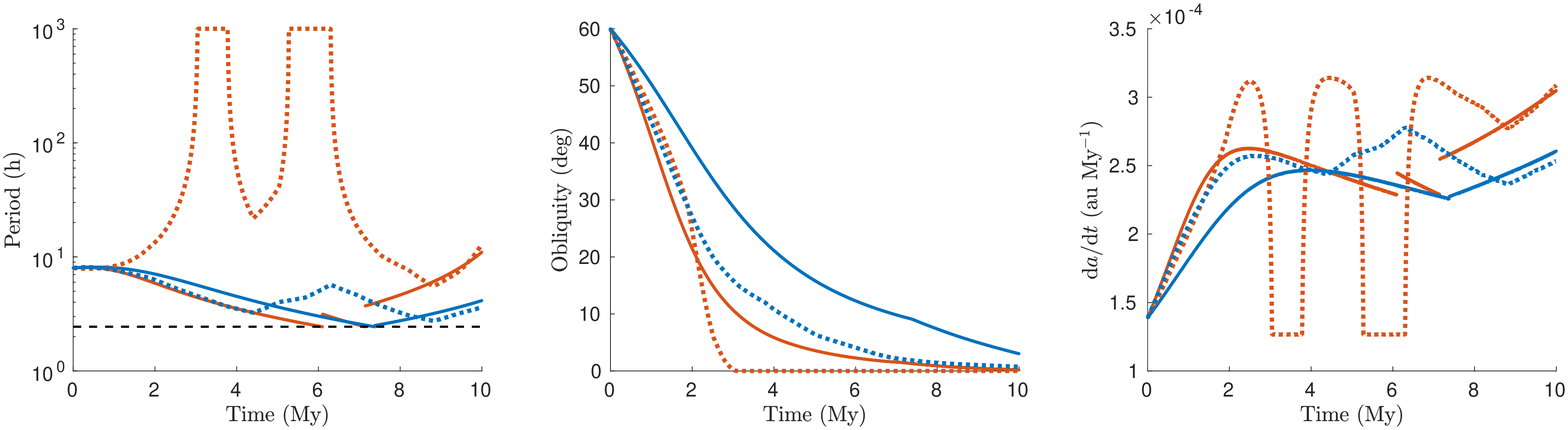}
    \caption{Evolution of rotation period $P$ (left panel), obliquity $\gamma$ (central
       panel), and semi-major axis drift $\text{d}a/\text{d}t$ (right panel), obtained
       with different integrators and different YORP models. Red color corresponds to
       the \textsc{mercury} integrator, while blue color to the \textsc{orbit9}
       integrator. Solid curves represent the simulations performed with the static YORP model,
       while dotted curves those with the stochastic YORP model.
       The diameter of the asteroid was assumed to be $D =
       2$ km. The black dashed line in the plots of the first column indicates the
       critical rotation period. The top row refers to results obtained for conductivity
       $K = 0.001$ W m$^{-1}$ K$^{-1}$, while the bottom row to $K = 0.01$ W m$^{-1}$
       K$^{-1}$.}
    \label{fig:mercury_orbit9_statstoc}
\end{figure*}
We performed a second test to understand how the static and the stochastic YORP models behave. We used the same set of 6 asteroids used in Sec.~\ref{ss:test1}, with same initial conditions and physical parameters. We integrated the orbits for 10 My with both packages, one time using the static YORP model of Sec.~\ref{s:yorp}, and another time using the stochastic YORP model of Sec.~\ref{s:stoc_yorp}. 

Figure~\ref{fig:mercury_orbit9_statstoc} shows the evolution of the rotation period $P$, the obliquity $\gamma$, and the semi-major axis drift $\text{d}a/\text{d}t$ obtained for the objects with diameter $D = 2$ km. 

The top row corresponds to the results obtained for $K = 0.001$ W m$^{-1}$ K$^{-1}$. Here,
it can be noted that the two objects integrated with the static YORP model (solid
curves) both undergo collisional re-orientation when the period becomes large, hence the
two packages behave correctly in this respect. When a collisional re-orientation occurs
the obliquity value changes randomly, and it can happen that the semi-major axis drift
changes the sign (see the red solid curve in the top right panel), thus decreasing the overall
effect of the Yarkovsky drift. 
Note that the timescale of the YORP cycle for these two objects is about 3-4 My.
On the other hand, the integrations obtained with the stochastic YORP model (dotted
curves) both present other interesting features. In the evolution obtained with
\textsc{mercury} (red dotted curve), the obliquity $\gamma$ inverts the trend two times by
effect of the occurrence of stochastic events that change the sign of the torque $g$.
However, in the overall evolution $\gamma$ is slowly approaching 0 deg. The rotation
period $P$ follows a random-walk, and it does not exceeds 10 hours. This has the effect of
maximizing the semi-major axis drift  $\text{d}a/\text{d}t$ on the 10 My evolution, as can
be seen from the top right panel of Fig.~\ref{fig:mercury_orbit9_statstoc}. 
In the evolution obtained with \textsc{orbit9} (blue dotted curve), the obliquity $\gamma$
undergoes several inversions of trend that produce a random walk during the first 5 My,
resulting in an evolution narrowed to a limited range. These alternating random obliquity
inversions have been found also in the model by \citet{2015ApJ...803...25C}, and the
phenomenon is referred to as \textit{self-governing YORP}. After this period, $\gamma$
finally decreases toward 0 deg. The rotation period has a random-walk evolution as well,
and it becomes larger and larger near 10 My, going toward the completion of one YORP
cycle. In both cases, the YORP cycle timescale is comparable to the whole 10 My timespan,
that is significantly larger than the timescale obtained in the static YORP model.

The bottom row of Fig.~\ref{fig:mercury_orbit9_statstoc} shows the evolution of the same
object, obtained for a conductivity of $K = 0.01$ W m$^{-1}$ K$^{-1}$. In all the four
presented cases, the obliquity $\gamma$ approached the asymptotic value of 0 deg. Here, in
the simulations performed with the static YORP model (solid curves) the period is
decreasing, and both objects approach the critical spin limit of 2.44 h. When this
critical value is reached, a mass shedding event occurs, and indeed we can notice a sudden
change in the period. This happens with both \textsc{mercury} and \textsc{orbit9},
confirming that the two packages behave correctly during the spin-up events. Therefore,
the YORP cycle for these two evolution ends at fast rotation, and the timescale are
comparable to those obtained in the case $K = 0.001$ W m$^{-1}$ K$^{-1}$. 
In the \textsc{mercury} evolution obtained with the stochastic model (red dotted curve), the
object reaches the 1000 h upper limit for the period twice, while the obliquity reaches 0
deg already during the first spin-down event. However, no jumps in $P$ and $\gamma$ due to
collision re-orientation events are present. This is also an effect of the stochastic YORP
model. As the period is freezed to the upper limit value, a torque $f$ that is
negative at 0 deg is selected during a stochastic event. On the other hand, the torque $g$
cannot change sign in this conductivity case. Since the object already reached the extreme
obliquity value at the time the stochastic event occurred, the change of sign in $f$ makes
the period start evolving again, this time towards lower values. In the \textsc{orbit9}
integration obtained with the stochastic model (blue dotted curve), the obliquity is slowly
approaching 0 deg, while the rotation period randomly walks at values smaller than 10 h.
This has the effect of maintaining a fairly constant semi-major axis drift of $2.5$-$2.7\times 10^{-4}$ au My$^{-1}$ for a significant fraction of the evolution. Again, in both cases the timescale
of the YORP cycle is significantly larger than the one obtained in the static YORP model.

These examples already show that, as expected, the stochastic YORP model increases the
YORP cycle timescale, and that the semi-major axis drift could be significantly larger
than that in the static YORP model.

\subsection{Evolution of an asteroid population}
To better understand the difference in the outcome between the static and the stochastic
YORP models, we performed a test with a larger sample of asteroids. For this purpose, we
took the nominal orbit of asteroid (163) Erigone, which is the parent body of the
corresponding asteroid family. We used the
AstDyS\footnote{\url{https://newton.spacedys.com/astdys/}} service to get the orbit
determined at time 59200 MJD, and we propagated the dynamics of 120 clones for 100 My. We
assumed the same initial orbit for all the clones, changing only the physical and spin
parameters. We assumed a fixed value of $\rho = 1200$ kg m$^{-3}$ for the density, that is
compatible with carbonaceous C-type asteroids. We fixed the heat capacity to $C=800$ J
m$^{-1}$ K$^{-1}$, and the thermal conductivity to $K = 0.01$ W m$^{-1}$ K$^{-1}$. The
initial rotation period was fixed to 6 h, while the obliquity was randomly generated
according to a uniform distribution for $\cos\gamma$ between -1 and 1. On the other hand,
the diameter $D$ was randomly generated between 1 and 5 km. Numerical integrations
were performed with the MVS method for \textsc{mercury} and with the multistep
method for \textsc{orbit9}, using 5 days time step for the orbital dynamics and
50 yr time step for the spin-axis dynamics.

%

\begin{figure*}[!ht]
    \centering
    \includegraphics[width=0.98\textwidth]{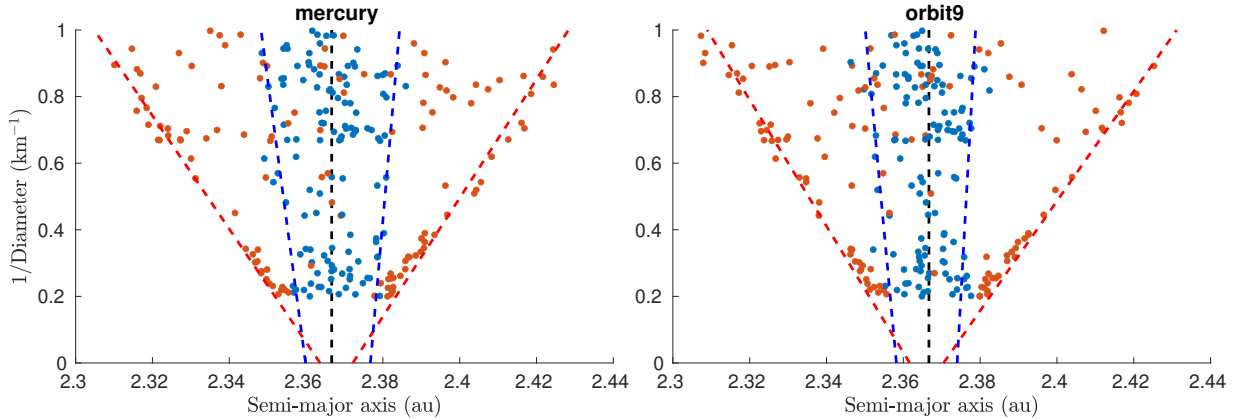}
    \caption{Final osculating semi-major axis vs. inverse of diameter of the 120 clones, for \textsc{mercury} (left panel) and \textsc{orbit9} (right panel). Blue dots are the results obtained with the static YORP model, while red dots are obtained with stochastic YORP model. The vertical dashed black line indicates the initial semi-major axis, while the dashed blue and red lines are the fits of the V-shapes.}
    \label{fig:dadt_mean}
\end{figure*}

\begin{table*}[!ht]
\centering
   \renewcommand{\arraystretch}{1.2}
\caption{Parameters of the V-shape fits.}
\label{tab:Vshape_fit}
\begin{tabular}{l|cc|cc|}
\cline{2-5}
                                       & \multicolumn{2}{c|}{\textsc{mercury}}                                  & \multicolumn{2}{c|}{\textsc{orbit9}}                                   \\ \hline
\multicolumn{1}{|l|}{V-shape side}     & \multicolumn{1}{c|}{$\alpha$}          & $\beta$                       & \multicolumn{1}{c|}{$\alpha$}           & $\beta$                      \\ \hline
\multicolumn{1}{|l|}{IN (stochastic)}  & \multicolumn{1}{c|}{2.3637$\pm$0.0023} & -0.0588$\pm$0.0035   & \multicolumn{1}{c|}{2.3617$\pm$0.0061} & -0.0527$\pm$0.0095 \\
\multicolumn{1}{|l|}{OUT (stochastic)} & \multicolumn{1}{c|}{2.3722$\pm$0.0056} & \hphantom{-}0.0562$\pm$0.0087 & \multicolumn{1}{c|}{2.3705$\pm$0.0023} & \hphantom{-}0.0607$\pm$0.0036 \\
\multicolumn{1}{|l|}{IN (static)}      & \multicolumn{1}{c|}{2.3599$\pm$0.0020} & -0.0116$\pm$0.0031            & \multicolumn{1}{c|}{2.3582$\pm$0.0039} & -0.0081$\pm$0.0061            \\
\multicolumn{1}{|l|}{OUT (static)}     & \multicolumn{1}{c|}{2.3767$\pm$0.0014} & \hphantom{-}0.0076$\pm$0.0022 & \multicolumn{1}{c|}{2.3740$\pm$0.0027} & \hphantom{-}0.0048$\pm$0.0041 \\ \hline
\end{tabular}
\end{table*}

We propagated the set of clones the first time with the static YORP model, and the second
time with the stochastic YORP model. Figure~\ref{fig:dadt_mean} shows the final
osculating semi-major axis versus the inverse of the diameter for both packages. From
these plots, it can be noted that the asteroids integrated with the static YORP model
(blue dots) are less dispersed from the initial semi-major axis value. Although an object
may be affected by a larger semi-major axis drift (compare for instance with
Fig.~\ref{fig:comparison_mercury_orbit9} and \ref{fig:mercury_orbit9_statstoc}) for a
while, the average drift is slowed down by the effect of collisional re-orientations
happening at long rotation period. When such events happen, the obliquity could
significantly change, making the drift $\text{d}a/\text{d}t$ suddenly change sign (see
also the solid red curve of the top right panel in
Fig.~\ref{fig:mercury_orbit9_statstoc}), hence decreasing the overall Yarkovsky effect.

The objects propagated using the stochastic YORP model (red dots) are much more
   dispersed in the semi-major axis. Moreover, even though this test does not represent an
   entirely consistent simulation of the evolution of an asteroids family, a strong
   correlation with the diameter can be noticed in the two panels of
   Fig.~\ref{fig:dadt_mean}, and the typical V-shape can be recognized. These results mean
   that the YORP cycle timescale is generally increased with the stochastic YORP model, as
   we have already seen in the examples of Fig.~\ref{fig:mercury_orbit9_statstoc}. Thus,
objects can maintain a faster Yarkovsky drift for longer, drifting further from the
initial semi-major axis value.

To further test whether the two packages provide the same statistical results, we
   fitted the edges of the V-shape. This was performed with a linear model of the form $a
   = \alpha + \beta D^{-1}$, using a least-square method described in
   \citet{2014Icar..239...46M}. The obtained numerical values are reported in
   Table~\ref{tab:Vshape_fit}. The fits of the V-shape for the simulations performed with
   \textsc{mercury} and \textsc{orbit9} agree within the error at 1-$\sigma$ level, for
both the static and the stochastic YORP models, providing evidence that the two packages
produce the same results, at least in the context of main-belt asteroids.

\section{SUMMARY AND CONCLUSIONS}
In this paper, we presented modified versions of the \textsc{mercury} and \textsc{orbfit}
numerical integration packages that take into account thermal effects acting on small
Solar System bodies. The Yarkovsky effect is computed with an analytical model and then
added to the gravitational vector field as a non-conservative force along the tangential
direction. The YORP effect governs the spin-axis dynamics of the object, and it is
integrated along with the orbital integration. We implemented a static and a stochastic
YORP model, which are both based on statistical studies performed by
\citet{2004Icar..172..526C}. Finally, we first provided some tests to show that the two
modified packages provide comparable results. Then we presented some simulations to show
how the two different YORP models behave. 

The modified versions of the \textsc{mercury} and \textsc{orbfit} packages developed in
this paper are publicly available. They are suitable for statistical studies of the
dynamics of large sets of asteroids.
For instance, these packages can be potentially used to investigate how the combined
Yarkovsky/YORP effects affect the dispersion and the age estimation of asteroid families
\citep[see e.g.][]{2020AJ....160..128M, 2021MNRAS.506.4302D}, or to clarify the origin and formation process of
the so-called \emph{YORP-eye} \citep{2016Icar..274..314P}. Additionally, it can be used to
test how the migration of small asteroids from the main-belt to the near-Earth region
changes with the introduction of the YORP effect in the dynamics
\citep{2017A&A...598A..52G}.

Finally, we point out that the solution adopted here for modeling the YORP effect is not
the only possible one, and the understanding of this thermal effect is still a matter of
research \citep[see][and references therein]{vokrouhlicky-etal_2015}. Other models have been
developed in recent years. For instance, \citet{2007AJ....134.1750N, 2008CeMDA.101...69S}
developed analytical models based on spherical harmonics, while
\citet{2012ApJ...752L..11G, 2017AJ....154..238G} found that the re-emission of the heat of
small boulders causes a torque component parallel to the surface, that is known as the
tangential YORP effect. When this component is added to the torques, new rotational
equilibria at which $P$ and $\gamma$ remain constant arise \citep{2019AJ....157..105G,
2021AJ....162....8G}. Although we designed the code so that changing the YORP effect model
is not technically complex, we leave these tasks for future developments of the software. 

\acknowledgements{
We thank Aldo Dell'Oro for the referee report that helped us to improve the paper.
The authors have been supported by the MSCA ETN Stardust-R, Grant
Agreement n. 813644 under the European Union H2020 research and innovation program. BN
also acknowledges support by the Ministry of Education, Science and Technological
Development of the Republic of Serbia, contract No. 451-03-68/2022-14/200104.}



\newcommand\eprint{in press }

\bibsep=0pt

\bibliographystyle{aa_url_saj}

{\small

\bibliography{sample_saj}
}

\begin{strip}

\end{strip}

\clearpage

{\ }

\clearpage

{\ }

\newpage

\begin{strip}

{\ }



\naslov{UPU{T}{S}TVO ZA AUTORE}


\authors{Marco Fenucci$^{1}$, Bojan Novakovi\'c$^{1}$}

\vskip3mm


\address{$^1$Department of Astronomy, Faculty of Mathematics,
University of Belgrade\break Studentski trg 16, 11000 Belgrade,
Serbia}


\Email{marco\_fenucci@matf.bg.ac.rs, bojan@matf.bg.ac.rs.rs}

\vskip3mm


\centerline{{\rrm UDK} \udc}


\vskip1mm

\centerline{\rit Orginalni nau\v cni rad}

\vskip.7cm

\baselineskip=3.8truemm

\begin{multicols}{2}

{
\rrm

TBA...

{\ }

}

\end{multicols}

\end{strip}


\end{document}